\documentclass[10pt, conference]{IEEEtran}
\usepackage{xcolor}
\usepackage{ifpdf}

\ifCLASSINFOpdf
  \usepackage[pdftex]{graphicx}
\else
\fi
\usepackage{amsmath}
\usepackage{algorithmic}

\usepackage{array}

\ifCLASSOPTIONcompsoc
 \usepackage[caption=false,font=normalsize,labelfont=sf,textfont=sf]{subfig}
\else
 \usepackage[caption=false,font=footnotesize]{subfig}

\usepackage{stfloats}
\usepackage{url}

\usepackage{booktabs}
\usepackage{multirow}
\usepackage{diagbox}
\usepackage{placeins}
\usepackage{balance}

\begin{document}
%

\title{A Comparative Study of Different Approaches for Tracking Communities in Evolving Social Networks}

\author{\IEEEauthorblockN{Ziwei He, Etienne Gael Tajeuna, Shengrui Wang}
\IEEEauthorblockA{Department of Computer Science\\
University of Sherbrooke\\
Sherbooke, QC, Canada\\
\{ziwei.he, etienne.gael.tajeuna, shengrui.wang\}@usherbrooke.ca}
\and
\IEEEauthorblockN{Mohamed Bouguessa}
\IEEEauthorblockA{Department of Computer Science\\
University of Quebec at Montreal\\
 Montreal, QC, Canada\\
bouguessa.mohamed@uqam.ca}
}


%


\maketitle

\begin{abstract}
In real-world social networks, there is an increasing interest in tracking the evolution of groups of users and detecting the various changes they are liable to undergo. Several approaches have been proposed for this. In studying these approaches, we observed that most of them use a two-stage process. In the first stage, they run an algorithm to identify groups of users at each timestamp. In the second stage, a pair-wise comparison based on a similarity measure is employed to track groups of users and detect changes they may undergo. While the majority of existing approaches use a two-stage process, they all run different algorithms to identify communities and rely on different similarity measures to track groups of users over time. Noting that the different approaches may perform differently depending on the dynamic social network under investigation, we decided to make a high level survey of some existing tracking approaches and then do a comparative analysis of some of them. In our analysis, we compared the algorithms in two main situations: $(1)$ when groups of users do not overlap and $(2)$ when the groups are overlapping. The study was done on three different testbeds extracted from the DBLP, Autonomous System (AS) and Yelp datasets.
\end{abstract}


%
\IEEEpeerreviewmaketitle

\section{Introduction}\label{sec:introduction}

Modeling and mining social networks has been a hot issue in the last decade, with many researchers seeking to reveal hidden patterns and their evolution. Traditional methods for analyzing social networks have focused on modeling the network as a static graph \cite{lancichinetti2008benchmark}, \cite{leskovec2010empirical}, where the behavior of the individuals is frozen in a snapshot. However, this type of modeling captures neither the temporal aspect nor the evolution of the network. Recent methods \cite{sarkar2005dynamic}, \cite{takaffoli2011tracking}, \cite{asur2009event}, model the graph as a series of frozen networks, where each network corresponds to a particular point in time. Such modeling has been useful to detect structural changes in the network \cite{takaffoli2011tracking}, \cite{asur2009event}, \cite{lee2014incremental}, \cite{tajeuna2015tracking}, \cite{bhat2015hoctracker} and to reveal important network information. To detect changes in dynamic social networks, two major approaches have been proposed. Some authors use a global approach \cite{sarkar2005dynamic}, \cite{rossi2013modeling}, \cite{skillicorn2013spectral}, in which the complete network is tracked over time to observe how nodes and edges behave. Others \cite{takaffoli2011tracking}, \cite{asur2009event}, \cite{tajeuna2015tracking}, \cite{bhat2015hoctracker} focus their efforts on tracking communities over time. In this paper, our focus is on the evolution of communities over time.

In the social network field, the term ``communities'', has been defined as groups of users in a way it can be taken as a partition of the social network, where each individual belongs to one and only one group at a time. In the real world, however, an individual may belong to several communities at the same time. In this last case, we talk about overlapping communities. 

Tracking the evolution of groups of users within social networks has attracted growing interest from researchers due to the wide variety of application domains, including the mining and analysis of sociological phenomena. For example, in criminology \cite{calvo2004social}, social network methodologies are used to discover and track groups of delinquent individuals over time in order to control them. In the public health field \cite{luke2007network}, social network strategies can be applied to discover the dynamics of certain subpopulations that are susceptible to a disease, or to predict the early stages of an epidemic.

In most existing work on analyzing the evolution of community structures in dynamic social networks, the important issues are how to discover transitions or critical events a community may undergo and how to track communities over time. Virtually all of the existing approaches \cite{takaffoli2011tracking}, \cite{asur2009event}, \cite{lee2014incremental}, \cite{tajeuna2015tracking}, \cite{bhat2015hoctracker}, \cite{greene2010tracking}, \cite{brodka2013ged} begin by considering the dynamic network under investigation as a series of static snapshot graphs at different time points. Then, using a community detection algorithm, they identify community structures at each of these snapshots independently. We note that in the existing approaches, the authors use different community detection algorithm to identify groups of nodes within the social network at each time-stamp. Moreover, they use different similarity measures to track and detect the changes the communities may undergo over time. Some authors, such as Brodka et al, in \cite{brodka2013ged} tested their approach with both overlapping and non-overlapping communities, while others tested on non-overlapping communities only. It is clear that the tracking results are highly dependent on the community detection method and the similarity measure used. The question that directly concerns us is, how these tracking algorithms are affected when using different similarity measures in cases where we have overlapping and non-overlapping communities.


\begin{figure}[!t]
\centering
{\includegraphics[width=3.5in]{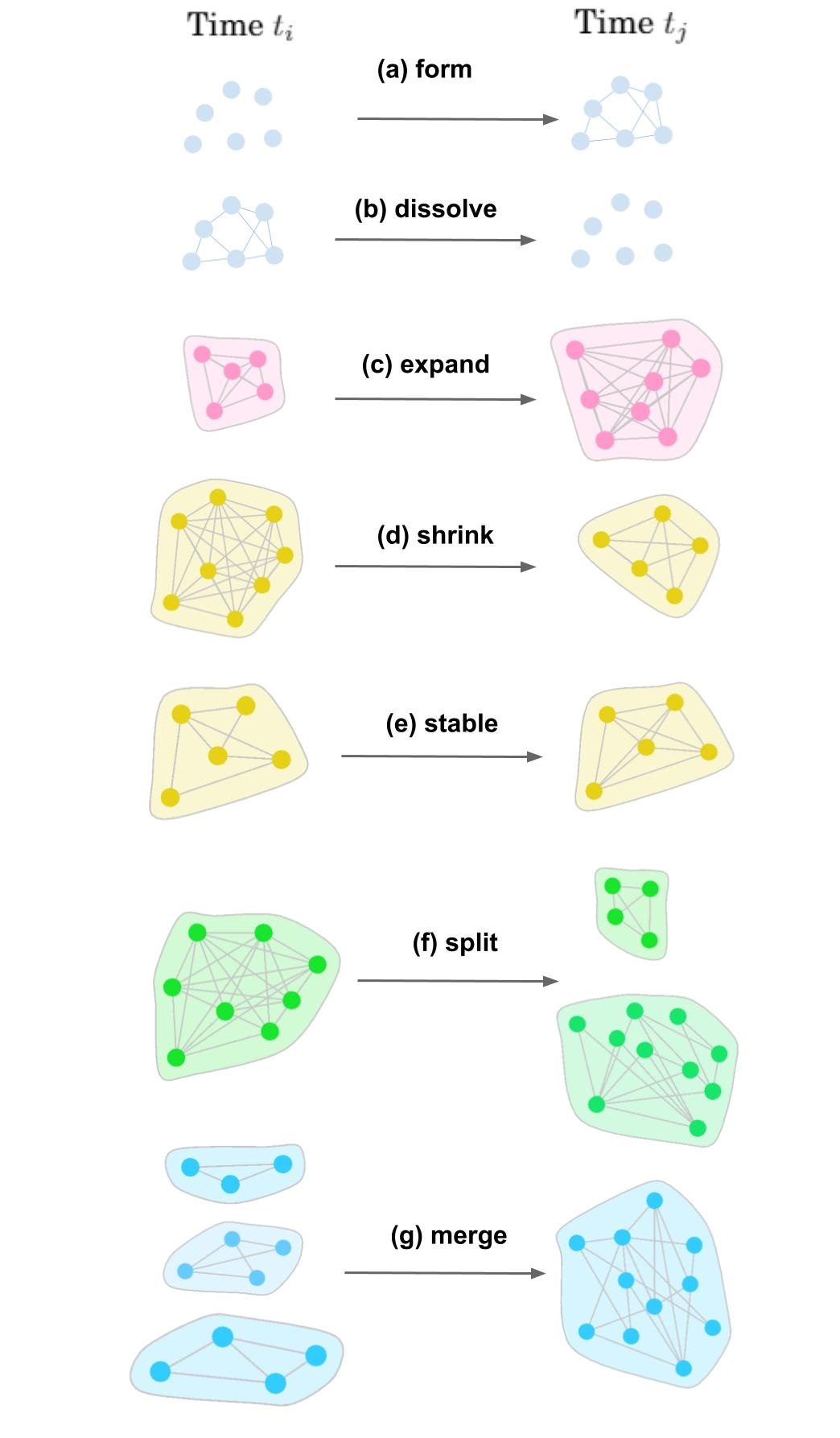}
\label{events}}
\caption{Possible events a community may undergo as time evolves from time $t_{i}$ to $t_{j}$}
\end{figure}

\section{Mainstream approaches for tracking communities}\label{algorithms}
Before talking about different algorithms used to track communities over time, we first define some important concepts.
\subsection{Concept definitions}\label{Definitions}
At any time $t_i$ we use the graph structure $g_{t_{i}}\; = \; (V_{t_{i}},E_{t_{i}})$ to represent the snapshot of the social network, where $V_{t_{i}}$ stands for the set of nodes and $E_{t_{i}}$ the set of edges. Hence, for a duration going from $t_1$ to $t_m$, we use the series $G\; =\; \left \{V_{t_{i}},\; E_{t_{i}}\; |\; 1\; \leq \; i \; \leq \; m \right \} \; = \; \left(g_{t_{i}}\right)_{1\; \leq \; i \; \leq \; m}$ to denote the time evolution of the social network. For each graph $g_{t_{i}}$, we define a set of subgraphs $\left\{C_{t_{i}}^{1},\; C_{t_{i}}^{2},\; ...,\; C_{t_{i}}^{q_{i}} \right\}$ representing the communities detected at time $t_{i}$ using a community detection algorithm. Each $C_{t_{i}}$ is a community of $G$, with $V_{t_{i}}$ and $E_{t_{i}}$ as its sets of nodes and edges, respectively. 

Given two communities $C_1$ and $C_2$, they are assumed to be similar when they respect a conditional criterion known as \textit{threshold}, keeping in mind that the notion of similarity varies from author to author. In our research, we found that most of the similarity measures used by the authors to track communities over time are based on the number of nodes shared. For example, the communities $C_1$ and $C_2$ can be considered similar if they have at least $30 \%$ of their nodes in common. Note that the \textit{threshold} criterion may be at the user’s discretion, or it may also be automatically set, as in \cite{takaffoli2011tracking}, \cite{tajeuna2015tracking}.

We denote by ``evolving community'' the sequence \cite{takaffoli2011tracking}, \cite{tajeuna2015tracking}, \cite{greene2010tracking}, \cite{brodka2013ged}, \cite{gliwa2013different} of communities tracked, where each community within this sequence indicates the status of the evolving community at a specific time point. For instance, $S_{C^a}\,=\,\{C_{t_1}^a,\,C_{t_2}^a,\,C_{t_3}^a,\,...,\,C_{t_9}^a\}$ will be the evolution of community $C^a$ from $t_1$ till $t_9$.
After identifying the communities over time, the authors in \cite{takaffoli2011tracking}, \cite{asur2009event}, \cite{lee2014incremental}, \cite{tajeuna2015tracking}, \cite{greene2010tracking},  \cite{brodka2013ged}  perform a pairwise comparison between the identified community structures. Here, we note that the comparison is performed by investigating the similarity of two communities identified at two consecutive or non-consecutive timestamps.

From one time to another, an evolving community may change its structure due to the arrival and/or departure of nodes and edges. Hence, given a community evolving in time, it may \textit{expand} because several nodes have joined the community, or \textit{shrink} because several nodes have left. In the same way, the community may \textit{split} into different communities, or several communities may \textit{merge} into one community. We may also observe a community that completely \textit{dissolves} over time. Fig. \ref{events} illustrates the different types of events a given community may undergo between two different times $t_i$ and $t_j$ ($t_i<t_j$). All of these changes are also known as critical events. 

\subsection{Algorithms}
\label{sec: algorithms}

\subsubsection{Asur et al. \cite{asur2009event}}
Asur et al. \cite{asur2009event} defined an event-based framework to characterize complex behavioral patterns of individuals and communities over time. In their paper, they first detected disjoint community structures for each snapshot of the network at a specific time using the Markov Clustering Algorithm \cite{tang2014detecting}. Then, they defined five events that communities can undergo between any two consecutive snapshots. The events are \textit{stable}, \textit{merge}, \textit{split}, \textit{form}, and \textit{dissolve}. To find these events, they compared two communities at two successive timestamps by investigating the number of nodes shared. These hypotheses for finding critical events were used to track the evolution of communities over time. 

Note that Asur et al. assumed a community discovered at given time $t_{i}$ to be dissolved if none of the nodes of this community was present in the network at the following time $t_{i+1}$. Therefore, given a community at time $t_1$, for example, if there is no community similar to it at time $t_2$, the community should stop being tracked. However, some nodes may be present at non-consecutive times, which means that a community may not \textit{dissolve} after only one time-step observation. \\

\subsubsection{Greene et al. \cite{greene2010tracking}}
Like Asur et al. in \cite{asur2009event}, Greene et al. adopted the same strategy to detect critical events the communities may undergo. Note that, in their experiments, rather than using the Markov clustering algorithm to identify the communities, Greene et al. used the Blondel modularity optimization algorithm \cite{blondel2008fast} to identify disjoint communities at different timestamps. 

Though the strategy used to identify critical events is similar to that in \cite{asur2009event}, there are differences in the definition of critical events. One such difference concerns the \textit{dissolve} event. Greene et al. assumed a community observed at a given time $t_{i}$ to be dissolved if, after $d\;>\;2$ consecutive times, none of its nodes is present in the graph. 

In their approach, Greene et al., identified sequences of communities that represent the evolution of community structures. To this end, at each time they compared the ratio of nodes shared among the communities at consecutive timestamps. Hence, two communities (identified at $t_i$ and $t_{i+1}$) are aligned in the same sequence if they share at least $k \%$ of nodes according to the Jaccard coefficient. Note that the condition they imposed for declaring a community dissolved allowed Greene et al. to discover evolving communities at non-consecutive timestamps. However, this evolving discontinuity is influenced by the user-determined parameter $d$.\\


\subsubsection{Takaffoli et al. \cite{takaffoli2011tracking}}
In their method, Takaffoli et al. used the local community mining algorithm \cite{chen2009local} to produce
sets of disjoint communities for each snapshot. In contrast to the previous approaches, in which most of the critical events took place only in consecutive timestamps, Takaffoli et al.'s method is capable of identifying critical events that occur at consecutive and non-consecutive timestamps. The authors can thus track communities evolving in s non-consecutive fashion. The tracking process operates by comparing the communities at different timestamps, using the following similarity measure:

\small
\begin{align}\label{modec sim}
Sim(C_{t_i},\; C_{t_j}) &= 
\begin{cases}
\frac{|V_{t_i}\; \bigcap \; V_{t_j}|}{max(|V_{t_i}|,\; |V_{t_j}|)} \;\; if & \frac{|V_{t_i}\; \bigcap \; V_{t_j}|}{max(|V_{t_i}|,\; |V_{t_j}|)}\; \geq \; k
\\
0 & otherwise
\end{cases}
\end{align}

\normalsize
It is important to note that the \textit{threshold} similarity $k$ is  automatically determined. The \textit{threshold} is evaluated using a text-mining approach. Therefore, the authors evaluate their approach on social networks that incorporate content information, such as DBLP and ENRON, where they can exploit the information shared between nodes.\\

\subsubsection{Brodka et al. \cite{brodka2013ged}}
In all of the above methods, the authors did not test their approaches on overlapping communities, which might reveal different information. To address this shortcoming, Brodka et al. developed a flexible approach called group evolution discovery (\textit{GED}) which is able to track disjoint and overlapping communities.

Note that in the previous approaches, to detect changes and track the evolution communities may undergo, authors relied on a measure based only on the proportion of nodes shared at two consecutive time. Brodka et al. extended this measure by including a topological metric in their comparison:

\small{
	\begin{align}\label{inclusion sim}
		I(C_{t_i},\; C_{t_{i+1}})\; = \; \frac{V_{t_i}\; \bigcap\; V_{t_{i+1}}}{V_{t_i}}\; \times \; \frac{\sum_{n\; \in \; V_{t_i}\; \bigcap\; V_{t_{i+1}}}\; NI_{C_{t_i}}(n)}{\sum_{n\; \in \; V_{t_i}}\; NI_{C_{t_i}}(n)}
	\end{align}
}
\normalsize

\noindent where $NI_{C_{t_i}}(n)$ reflects the importance of node $n$ within the community $C_{t_i}$. This measure might be any centrality metric (centrality, social position, degree, etc.). With the added topological metric, their comparison is also able to consider the inter-relation among the nodes in a community. 

Though the comparison is done at consecutive timestamps, the inclusion effect of their comparison metric helps Brodka et al’s method track overlapping and non-overlapping communities. It is important to note that in their experiments, they made use of the \textit{CPM} algorithm \cite{adamcsek2006cfinder} identify overlapping communities at each timestamp before the tracking.\\

\subsubsection{Gliwa et al. \cite{gliwa2013different}}
Like the \textit{GED} framework used in \cite{brodka2013ged}, Gliwa et al. proposed the Stable Group Changes Identification \textit{SGCI} method to track and predict community changes in dynamic social networks. Note that in their approach, they used the \textit{CPM} algorithm \cite{adamcsek2006cfinder} to extract overlapping communities at each timestamp of the dynamic social network. However, rather than using the inclusion metric (Eq.(\ref{inclusion sim})) as in \cite{brodka2013ged}, they proposed a modified Jaccard (\textit{MJ}), given as follows:

	\begin{align}\label{modified jac}
		MJ(C_1,C_2) &= max\left(\frac{C_1\bigcap C_2}{C_1},\,\frac{C_1\bigcap C_2}{C_2}\right)
	\end{align}

\noindent where two communities ($C_1$, $C_2$) are assumed to be similar if $MJ(C_1,C_2)\geq 0.5$.\\

\subsubsection{Tajeuna et al. \cite{tajeuna2015tracking}}

In order to track communities and extract critical events, Tajeuna et al. first represented each community as a vector (called the \textit{transition probability vector}) indicating the number of nodes shared by the communities over time. Then, they compared the corresponding vectors of different communities. Specifically, given two communities $C_{t_i}$ and $C_{t_j}$ with \textit{transition probability vectors} $v_i$ and $v_j$, respectively, they calculated the similarity between the two communities as

{
\begin{equation}
sim(C_{t_i},\; C_{t_j})=\begin{cases}
\sum_{\alpha=1}^{N_{c}}\; 2\;\frac{p_{i,\; \alpha}\; \times\; p_{j,\; \alpha}}{p_{i,\; \alpha}\; +\; p_{j,\; \alpha}}\;\; 
\\\\
\;\;\; if\;\; \sum_{\alpha=1}^{N_{c}}\; 2\;\frac{p_{i,\; \alpha}\; \times\; p_{j,\; \alpha}}{p_{i,\; \alpha}\; +\; p_{j,\; \alpha}}\; > \;\lambda\\\\
0\;\;\; otherwise
\end{cases}
\label{similarity}
\end{equation}
}

\noindent where $\lambda$ is the junction point between the two Gamma curves estimated from the non-zero values obtained when scoring the similarity between two transition probability vectors; $p_{i,\; \alpha}$ and $p_{j,\; \alpha}$ are the respective components of vectors $v_i$ and $v_j$.

Using the similarity measure described in Eq.(\ref{similarity}), they defined the evolution of community $C_{t_i}$ as the sequence of sorted communities $S_{C_{t_i}}\; = \; \{C_{t_i},\,C_{t_{i\; + \; \eta}},\,...,\,C_{t_k}\}$, $t_i\; < \; t_k \; \leq \; t_m$ such that all communities in $S_{C_{t_i}}$ are similar. \\

\subsubsection{Summary}
Table \ref{overview} presents a summary of the mainstream approaches. The first column identifies the different approaches. Column two indicates the community detection algorithm used by each, and column three, the type of communities identified by the particular community detection algorithm. The fourth column indicates the similarity measure used to compare the communities at different timestamps. In the fifth column, the strategy used to set the similarity threshold is identified. Finally, the last column presents the tracking result, which specifies whether the approach can track communities evolving in a consecutive or non-consecutive way.

It is worth noting that the notion of non-consecutive evolving communities taken in ~\cite{greene2010tracking} differs slightly from the one in ~\cite{takaffoli2011tracking} and ~\cite{tajeuna2015tracking}. For instance, say we assume with Greene et al.~\cite{greene2010tracking} that a community dissolves if no observation of it is found after $d = 2$ consecutive timestamps. It will thus be impossible to recognize it if it later exists, rendering their approach unable to discover communities that evolve non-consecutively, with an interval of $d > 2$ consecutive timestamps during which there is no observation. Whereas in ~\cite{takaffoli2011tracking} and ~\cite{tajeuna2015tracking}, a community is assumed to be dissolved if there is no observation of it at any time until the last time-stamp observation.

\begin{table*}[!t]
\centering
\caption{Overview of the mainstream approaches.}
\label{overview}
\begin{tabular}{@{}cccccc@{}}
\toprule
\textbf{Approach}  & \textbf{Community detection} & \textbf{Type of communities} & \textbf{Similarity measure} & \textbf{Threshold setting} & \textbf{Tracking}\\
\midrule
Asur et al. \cite{asur2009event} & Markov clustering \cite{tang2014detecting} & Disjoint & Modified Jaccard & Manually set & Consecutive\\
\midrule
Greene et al. \cite{greene2010tracking} & Blondel modularity \cite{blondel2008fast} & Disjoint & Jaccard coefficient & Manually set & Consecutive \& non-consecutive \\
\midrule 
Takaffoli et al. \cite{takaffoli2011tracking} & Local community mining \cite{chen2009local} & Disjoint & Eq.(\ref{modec sim}) & Automatically set & Consecutive \& non-consecutive \\
\midrule 
Brodka et al. \cite{brodka2013ged} & \textit{CPM} \cite{adamcsek2006cfinder} & Overlapped & Eq.(\ref{inclusion sim}) & Manually set & Consecutive\\
\midrule
Gliwa et al. \cite{gliwa2013different} & \textit{CPM} \cite{adamcsek2006cfinder} & Overlapped & Eq.(\ref{modified jac}) & Manually set & Consecutive \\
\midrule
Tajeuna et al. \cite{tajeuna2015tracking} & Infomap \cite{rosvall2008maps} & Disjoint & Eq.(\ref{similarity}) & Automatically set & Consecutive \& non-consecutive\\
\bottomrule
\end{tabular}
\end{table*}

Though the community detection algorithms and similarity measures for tracking communities are different, there are several resemblances in some of the approaches. For instance, the tracking principles used in ~\cite{gliwa2013different} and ~\cite{brodka2013ged} are equivalents, differing only in the similarity measure used. Moreover, Gliwa et al. ~\cite{gliwa2013different} already made the comparison of their approach with that of Brodka et al.~\cite{brodka2013ged}. Therefore, rather than repeating both of these approaches, we will instead run the approach in ~\cite{brodka2013ged} with the one in ~\cite{greene2010tracking}, where the Jaccard coefficient is fully used.

In the work done by Asur et al.~\cite{asur2009event}, no community tracking process was specified. All that was demonstrated in their work is how to discover critical events that an evolving community may undergo at consecutive timestamps. In other words, they developed an algorithm for tracking events over time. However, since the focus of this paper is not to detail the various methodologies used to identify critical events, we have decided to avoid implementing the approach in ~\cite{asur2009event} which is hard to compare with others returning sequences of communities.

For the above reasons, then, we will only present experimental comparisons of the approaches in  ~\cite{takaffoli2011tracking}, ~\cite{tajeuna2015tracking}, ~\cite{greene2010tracking} and ~\cite{brodka2013ged}.\\


\section{Experiments}\label{section: first}

\subsection{Data Description}


In this section we validate four algorithms on three real-world datasets: the seventh version of the DBLP dataset\footnote{$http://arnetminer.org/citation$}, the Autonomous Systems (AS) dataset\footnote{$http://snap.stanford.edu/data/as.html$} and the YELP dataset\footnote{$http://www.yelp.ca/academic\_dataset$}. 

\begin{itemize}
\item The DBLP dataset contains co-publications of authors. For each published paper, it contains the paper's title, the authors, the year, the publication venue, the index identification of the paper and the identifications of references to the paper. We built undirected, unweighted graphs between co-authors and cited authors in the fields of data mining and artificial intelligence from 2011 to 2016, taking each year as a snapshot. Authors are represented by nodes, and co-authorships and paper citations by edges.
\item The AS dataset contains the daily communication network of who\-talks\-to\-whom from the Border Gateway Protocol logs. We built undirected, unweighted graphs on communication networks on a daily basis from 3 October 1999 to 2 January 2000, where each identifier is considered as a node and a relation between two identifiers is taken as an edge.
\item In the YELP dataset, there are three main objects: “Business”, “Review” and “User”, giving information on businesses reviewed by users. We focused on the “User” object: for each user having friend(s), we created undirected, unweighted edge(s) between this user and his or her friend(s). The graph construction is done on a monthly basis from August 2009 to July 2014.
\end{itemize}

Several methods exist for detecting communities in social networks. In this paper we use two algorithms: \textit{Infomap}~\cite{rosvall2008maps} (to detect non-overlapping communities) and \textit{CPM}~\cite{fortunato2010community} (to detect overlapping communities). These two detection methods can detect very different numbers of communities in the three datasets we are going to demonstrate on, so for the purpose of keeping appropriate numbers of communities for us to track regardless of whether the communities are overlapping or disjoint, we chose different numbers of snapshots on which to apply \textit{Infomap} and \textit{CPM}. Table ~\ref{tab:ds} contains the descriptions of the datasets used for testing overlapping and disjoint communities, respectively. The second and third columns present the total number of nodes per social network. The fourth and fifth columns show the number of snapshots we chose for each social network.\\

\begin{table}[!t]
    \centering
    \caption{Dataset Descriptions Overall.}
    \label{tab:ds}
    \begin{tabular}{@{} ccccc @{}}
\toprule
     & \multicolumn{2}{c}{Nodes} & \multicolumn{2}{c}{Snapshots} \\ \midrule
\#Network & Disjoint     & Overlapping    & Disjoint       & Overlapping      \\ \midrule
DBLP & 181474       & 181474     & 6              & 6            \\ \midrule
AS   & 6505         & 6741       & 20             & 60           \\ \midrule
YELP & 34276        & 3904     & 20             & 118           \\ \bottomrule 
\end{tabular}
\end{table}


\subsection{Procedure}

As in the existing approaches described in section~\ref{sec: algorithms}, we adopted a two-step approach.  We first identify overlapping communities using the \textit{CPM} algorithm and non-overlapping communities using the Infomap algorithm. The different algorithms are then compared by calculating the scores of the Jaccard coefficient between pairs of communities at distinct timestamps. For the metrics given in Eq.(\ref{modec sim}), Eq.(\ref{inclusion sim}) and Eq.(\ref{similarity}) we compare the pairs of communities at distinct timestamps as well.

Using a mixture of two Gaussian probability density functions, we automatically extract \cite{tajeuna2015tracking} the optimal threshold that characterizes the suitable similarity between two communities in the overall dataset under investigation. Note that this is done separately for each of the types of scores obtained. Thus, for each score obtained by a metric, we automatically identify the suitable threshold.

After determining the different thresholds, we run the different algorithms in \cite{takaffoli2011tracking}, \cite{tajeuna2015tracking},  \cite{greene2010tracking} and \cite{brodka2013ged} and  to identify the sequences of communities that reflect the evolution of communities over time. With the use of two general criteria (described below), we evaluate the purity of the evolving communities obtained in the four different cases.

In summary, the following steps can be processed to obtain the results given in detail in the next subsections:
\begin{itemize}
\item \textbf{Step 1}: Using the \textit{CPM} and \textit{Infomap} algorithms, identify the overlapping and non-overlapping communities at each timestamp.
\item \textbf{Step 2}: Using the Jaccard coefficient as given in \cite{greene2010tracking}, calculate the score for pairs of communities at distinct timestamps. Repeat the process for the metrics given in Eq.(\ref{modec sim}),  Eq.(\ref{inclusion sim}) and Eq.(\ref{similarity}).
\item \textbf{Step 3}: For each of the scores obtained in \textbf{Step 2}, automatically identify the suitable threshold using a mixture of Gaussians, as explained in \cite{tajeuna2015tracking}.
\item \textbf{Step 4}: With the thresholds obtained in \textbf{Step 3}, run the corresponding algorithm given in \cite{brodka2013ged}, \cite{greene2010tracking}, \cite{tajeuna2015tracking} and \cite{takaffoli2011tracking} to track the communities over time.
\item \textbf{Step 5}: Calculate the purity (explained below) of the sequences obtained with all the different approaches.\\
\end{itemize}

\subsection{Similarity Threshold Selection}
\label{section: threshold}
\begin{figure*}[!t]
    \centering
    \subfloat[YELP\_Jaccard.]{\includegraphics[width=0.3\linewidth]{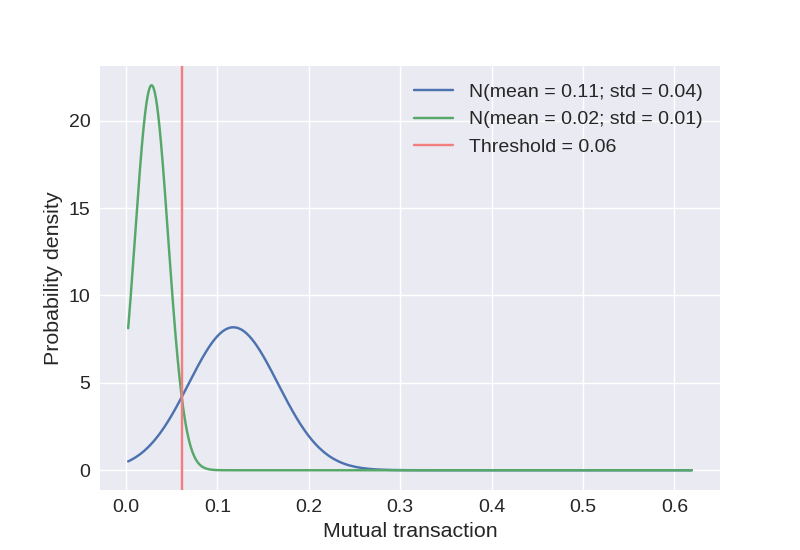}\label{fig:a}}\quad
    \subfloat[DBLP\_Jaccard.]{\includegraphics[width=0.3\linewidth]{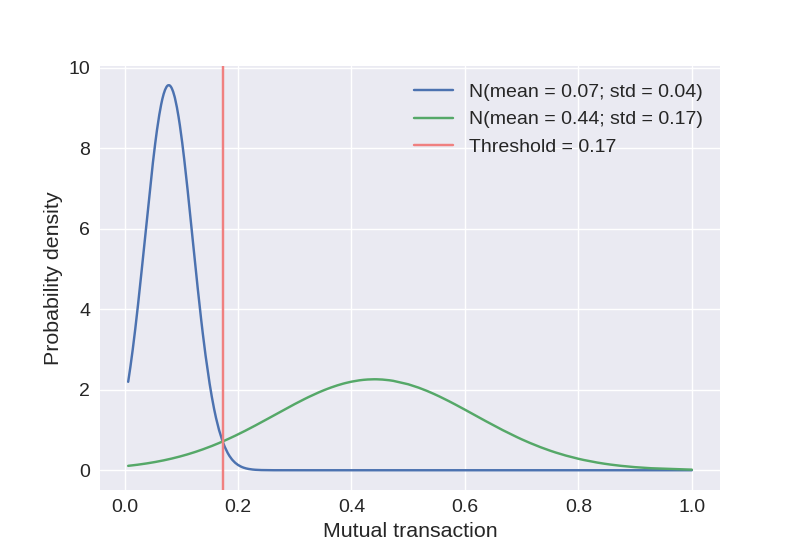}\label{fig:b}}\quad
    \subfloat[AS\_Jaccard.]{\includegraphics[width=0.3\linewidth]{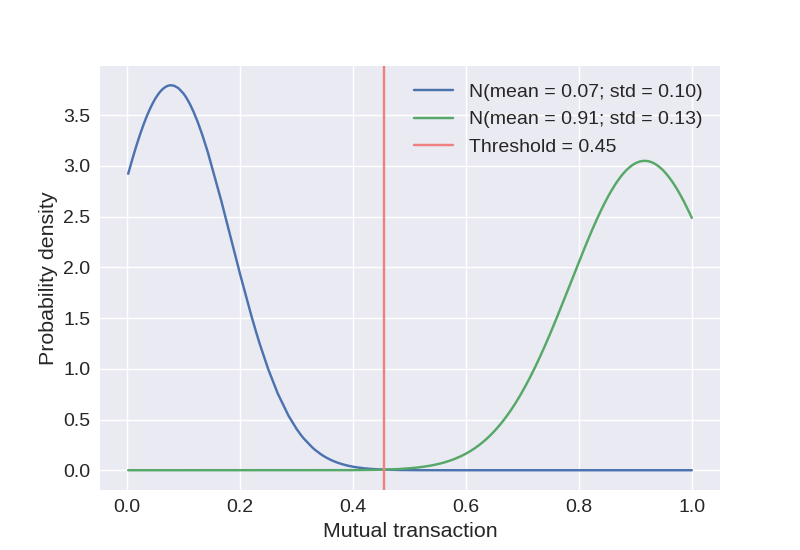}\label{fig:c}}\quad
    \medskip
    \subfloat[YELP\_Mutual.]{\includegraphics[width=0.3\linewidth]{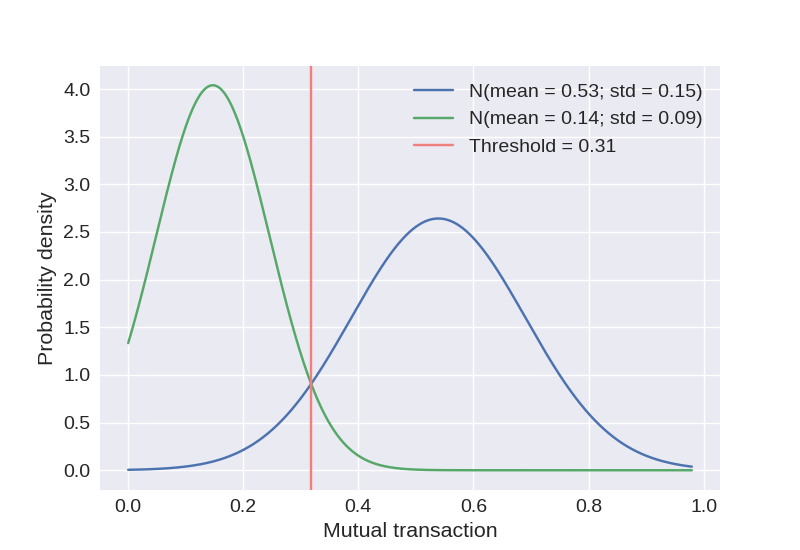}\label{fig:d}}\quad
    \subfloat[DBLP\_Mutual.]{\includegraphics[width=0.3\linewidth]{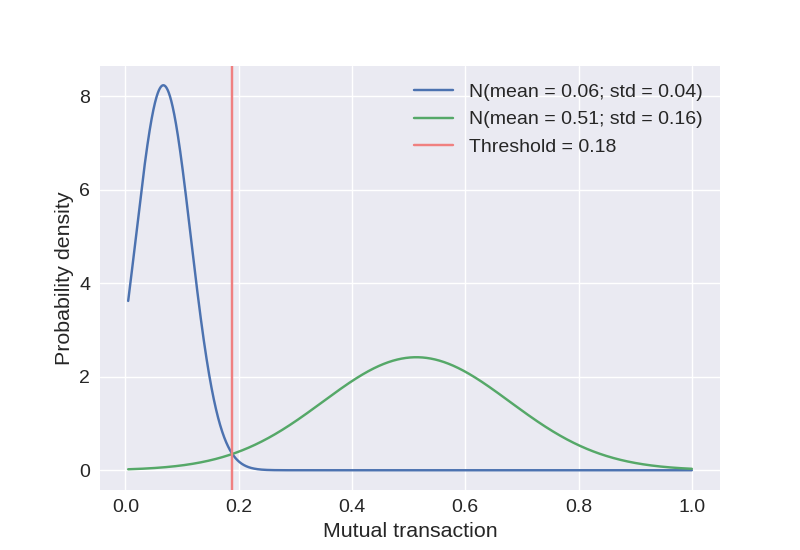}\label{fig:e}}\quad
    \subfloat[AS\_Mutual.]{\includegraphics[width=0.3\linewidth]{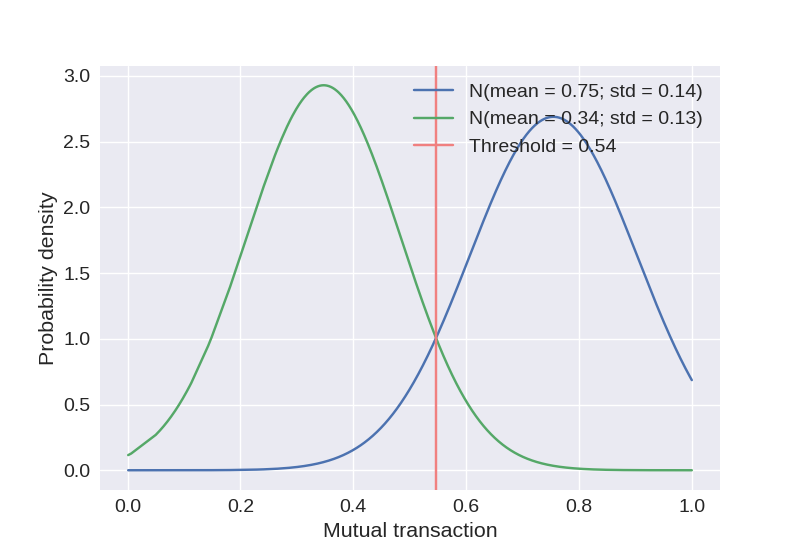}\label{fig:f}}
    \caption{Threshold of similarity as the junction point of two Gaussian density curves.}
\label{fig: threshold}
\end{figure*}

To determine the threshold that yields the best similarity the similarity between two detected communities, we look at the mixture of two Gaussians applied on the distribution of various similarities in all datasets. As an example, consider the Jaccard similarity~\cite{jaccard1912distribution} (used in the methods of Greene et al. and Tajeuna et al.) and the mutual similarity (used in the method of Tajeuna et al.) implemented on overlapping communities from all the three datasets. From the two Gaussians applied on the various distributions, we select as similarity threshold the mutual transition where the two Gaussians meet. The curves in Fig~\ref{fig:b} and Fig~\ref{fig:e} lie within $[0.15, 0.2]$, which indicates that the communities in DBLP change relatively quickly over time. On the other hand, the AS network is obviously more stable because the curve lies within $[0.4, 0.6]$, which indicates that the evolving communities extracted here are more "alike". This is shown in Fig~\ref{fig:c} and Fig~\ref{fig:f}. 
Below, we use the similarity threshold selecting using the mixture of two Gaussians to compare the different approaches.\\

\subsection{Evaluation} 

For an objective comparison of the evolving communities (sequence of communities) obtained by different approaches, we look at the tracking results with respect to the original community of each evolution sequence. This means we only compare the sequences starting with the same community. The comparison procedure is as follows: First, we adopt a general criterion based on the resemblance between each pair of selected communities in the evolving communities. The resemblance between two communities $C_i$ and $C_j$ is evaluated by the popular Pearson correlation coefficient defined as follows:
\footnote{Note that all values of the Pearson correlation coefficient are normalized within $[0,1]$}.
\begin{equation}
    \label{equation:pearson1}
    \rho _{C_{i}, C_{j}} = \frac{\left ( v_{i} - \bar{v_{i}} \right )\cdot \left ( v_{j} - \bar{v_{j}}\right )}{\left \| \left ( v_{i} - \bar{v_{i}} \right ) \right \|\cdot \left \| \left ( v_{j} - \bar{v_{j}}\right ) \right \|}, 
\end{equation}

\noindent where $v_{i}$ and $v_{j}$ are the corresponding transition probability vectors of communities $C_{i}$ and $C_{j}$, respectively. They reflect the number of shared nodes between $C_i$ ($C_j$) and each of the remaining communities discovered over the whole time interval. $\bar{v_{i}}$ and $\bar{v_{j}}$ are their respective mean values. From equation \ref{equation:pearson1}, we calculate the global resemblance of an evolving community $S_{C} = C_{a} \rightarrow C_{t_{a + \eta }} \rightarrow ... \rightarrow  C_{b}$ as the Average Pearson Correlation Coefficient (APCC), given as follows:
\begin{equation}
    \label{equation:pearson2}
    Avg\left ( S_{C} \right ) = \sum_{C_{i}\in S_{C}}\cdot \sum_{C_{j}\in S_{C}} \rho  _{C_{i}, C_{j}}
\end{equation}

As another criterion of resemblance, we look at the Average Proportion of Nodes Persisting (APNP) in an evolving community $S_{V_{C}}$, expressed as follows:

\begin{equation}
    \label{equation:np}
    N_{p}(S_{V_{C}}) =   \frac{\sum_{\eta = 1}^{b - a}\left |  V_{C_{a}}\bigcap V_{C_{a + \eta }}\right |}{\left |  V_{C_{a}}\right |} 
\end{equation}

\noindent where $S_{V_{C}} = \left \{ V_{C_{a}}, V_{C_{a + \eta }}, ..., V_{C_{b}} \right \}$ is the set of nodes corresponding to the sequence of community $（S_{C}$. Again, in the experiment the values have been normalized.\\

Evolutions of communities over time comprise what we refer to as community sequences. Note that different approaches could track community sequences in a different way. We compare how different methods track community sequences by considering the APCC and APNP values for each tracking method on each community sequence.\\

\subsection{Experiment Based on Overlapping Communities Extracted by CPM}

As a method for group extraction, \textit{CPM} was utilized. The groups were detected for $k = 4$, for undirected and unweighted social networks. Table ~\ref{tab:cpm} shows the description of the processed data: The first column shows the average number of communities detected by the \textit{CPM} algorithm per snapshot, while the second column shows the average size of the groups per snapshot.
\begin{table}[!t]
    \centering
    \caption{Data Descriptions for Overlapping Communities.}
    \begin{tabular}{{@{}ccc@{}}}
        \toprule
         Network    & \#avg\_com      &   \#com\_size    \\\midrule
         DBLP       &     1983        &       8          \\
         AS         &     30           &     40           \\
         YELP       &     5           &     　20         \\\bottomrule
    \end{tabular}
    \label{tab:cpm}
\end{table}

Table~\ref{tab:threshold_o} gives the similarity threshold values for each of the four tracking methods. \textit{Jaccard}, \textit{Modec}, \textit{Inclusion} and \textit{Mutual} are the different similarity measures considered for these tracking methods. Note that for the similarity measure \textit{Inclusion} given by Brodka et al.~\cite{brodka2013ged}, for each pair of communities, we calculated both $I(C_i, C_j)$ and $I(C_j, C_i)$; this explains why 2 similarity threshold values are listed in the fourth column.\\

\begin{table}[t]
\centering
\caption{Similarity Threshold for Each Method on Overlapping Communities. Greene et al. using Jaccard coefficient, Takaffoli et al. using Modec similarity, Brodka et al. using the Inclusion similarity and Tajeuna et al. Mutual transition.}
\label{tab:threshold_o}
\begin{tabular}{@{}ccccc@{}}
\toprule
& Greene et al.  & Takaffoli et al.  & Brodka et al.  & Tajeuna et al.  \\ \midrule

DBLP &     0.17  &     0.24      &       0.15 / 0.11   &         0.19       \\\midrule
AS   &        0.45  &      0.39      &      0.40 / 0.41       &        0.55      \\ \midrule
YELP &    0.06    &     0.08         &         0.10 / 0.03        &    0.32                   \\\bottomrule
\end{tabular}
\end{table}

\subsubsection{Quantity of Tracking}
\label{quatity_o}

\begin{figure}[t]
	\centering
    \caption{Tracking Quantities on Overlapping Communities.}
    \label{fig: quantity_o}
    \includegraphics[width=0.5\textwidth]{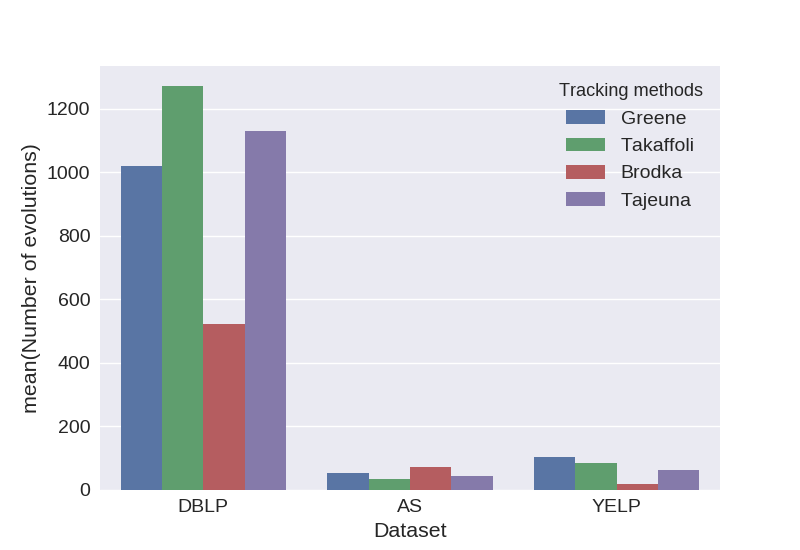}
\end{figure}

After implementing all four tracking approaches over these data streams which contain overlapping communities, we observed that a number of communities undergo evolutions over time; we refer to these as evolving communities. Fig~\ref{fig: quantity_o} shows the numbers of evolving communities found by different approaches on each dataset. We can then draw a number of conclusions on how these approaches compare with respect to the number of evolutions they find. 

We look at Fig~\ref{fig: quantity_o}, X-axis represents different datasets, they are DBLP, AS, YELP separately. Y-axis indicates the number of evolving communities each approach find. Different colors of the bar represents different approaches. Because different dataset has different size, it leads to very different number of communities will be found, in our case, DBLP dataset is much more larger than AS and YELP, so, the first five columns corresponded to DBLP have the higher values than other columns.

From the first five columns we observe that, for datasets with large graphs and small average size of overlapping communities like DBLP, the approaches of Greene et al., Takaffoli et al. and Tajeuna et al. can find sufficient numbers of evolving communities, while that of Brodka et al. can only find half of them.

As a middle point, YELP includes overlapping communities of medium average size and medium-size graphs, which is the most common situation in real life. For this dataset, only the approach of Brodka et al. doesn't find enough evolving communities, and others perform good. Therefore, we can say that Brodka’s approach has less chance of receiving a good audience in tracking small-size to medium-size communities when the quantity of tracking is of interest.

We also noticed that, for datasets which include large average size of overlapping communities such as AS, all the approaches perform well, especially approach Brodka et al..

Going through all the results, we can conclude that generally, the approaches of Greene et al., approach of Tajeuna et al. and Takaffoli et al. are efficient enough to capture most potential evolutions over time. Approach of Brodka et al. only performs well on big community-size dataset.\\

\subsubsection{Quality of Tracking}
\label{quality_o}

\begin{figure*}[!t]
    \centering
    \subfloat[DBLP\_APCC.]{\includegraphics[width=0.4\linewidth]{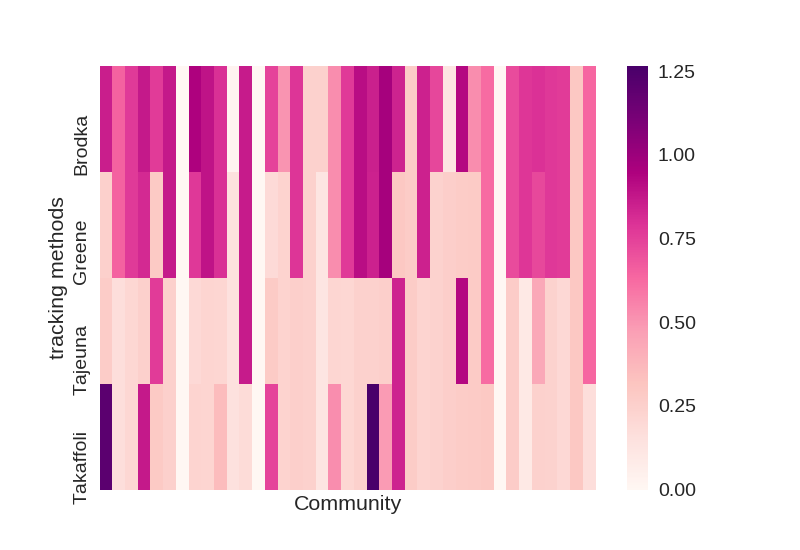}\label{fig:o_b}}\quad
    \subfloat[AS\_APCC.]{\includegraphics[width=0.4\linewidth]{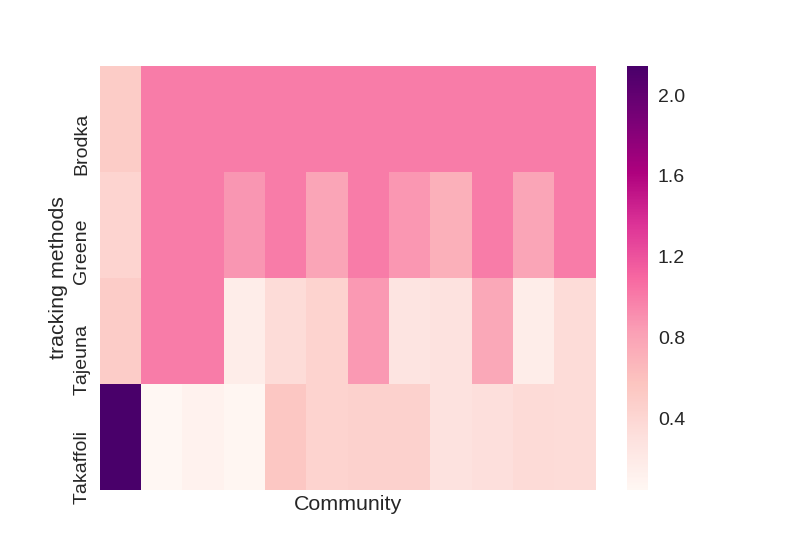}\label{fig:o_c}}\quad
    \medskip
    \subfloat[DBLP\_APNP.]{\includegraphics[width=0.4\linewidth]{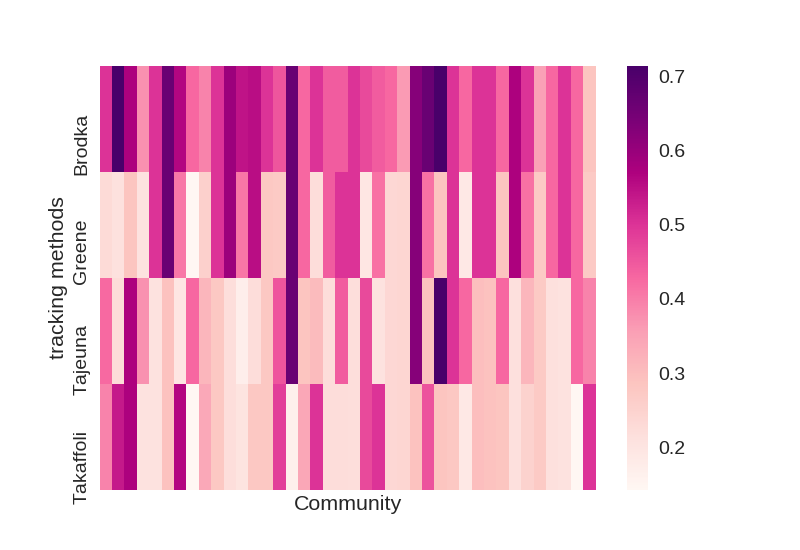}\label{fig:o_e}}\quad
    \subfloat[AS\_APNP.]{\includegraphics[width=0.4\linewidth]{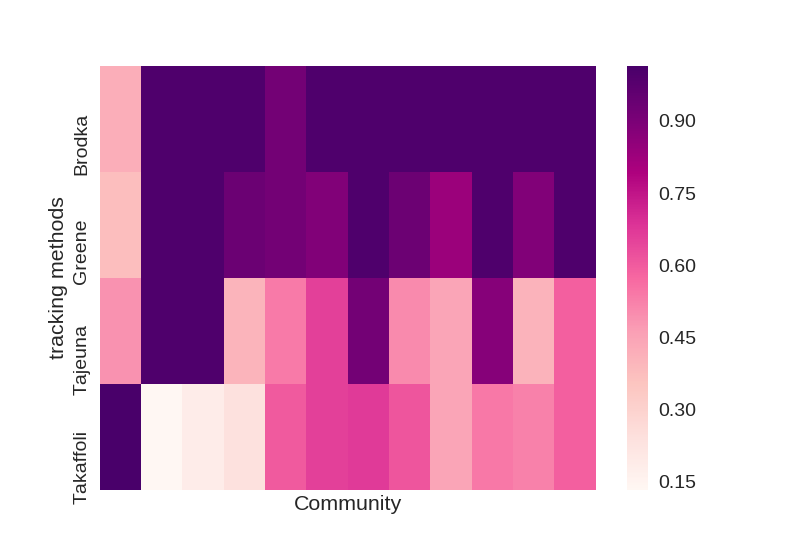}\label{fig:o_f}}
    \caption{In this heatmap, the darker the color of the cell, the higher the value. For (a) and (c) we removed the communities which all four approaches can track well (APCC and APNP are all above $0.3$) and for (b) and (d) we again removed the communities which all four approaches track well (APCC and APNP are all above $0.8$).}
\label{fig: evaluation_o}
\end{figure*}

From the previous section (Sec~\ref{quatity_o}), we know different approaches are capable of tracking different communities or, we could say, different numbers of communities. Still, as mentioned before, the variety of tracking algorithms means that they could track the same community differently. In Fig~\ref{fig: evaluation_o}, we show the quality of tracking of the four tracking algorithms. In each sub-figure, the X-axis presents those communities that all the approaches could successfully track and the Y-axis indicates the quality of the tracking algorithms on those communities. If we consider columns, each column has four values corresponding to how one community has been tracked by the four approaches. Each row presents the quality of a given tracking method over all the communities. Taking this into account, it is simple to estimate which approach has better results by looking for the addresses of darker-colored cells.

For this purpose, we look at the global performance using the statistical criteria defined in formulas~\ref{equation:pearson2} and \ref{equation:np} to define the quality of tracking, and make the comparison by looking at two specific criteria.

The numbers of communities selected as a starting point for an evolution which all the approaches can successfully track, used to plot the heatmap, are as follows: 
\begin{itemize}
\item For DBLP, there are 440 communities.
\item For AS, there are 19 communities.
\item For YELP, there is only 1 community.
\end{itemize}

For the YELP dataset, due to the limited number of overlapping communities, there exists only one community that all four approaches can successfully track. Therefore, for this section only the demonstration on the DBLP and AS datasets is considered.

Fig~\ref{fig: evaluation_o} gives the heatmap for APCC and APNP. In Fig~\ref{fig:o_b} and Fig~\ref{fig:o_c}, each column illustrates the four APCC values for the evolutions of the same community tracked by the Brodka et al., Greene et al., Tajeuna et al. and Takaffoli et al. approaches. Fig~\ref{fig:o_e} and Fig~\ref{fig:o_f} presents the same illustration for APNP.


As seen in Fig~\ref{fig:o_b} and Fig~\ref{fig:o_e}, the dataset DBLR contains big graphs and a large number of communities, so all the tracking approaches can successfully track a lot of communities. Looking at them, we observe that the darkest cubes appear in the row corresponding to the approach of Brodka et al., and Greene et al. also performs well. Hence, we can conclude temporarily that, for datasets with small-size communities like DBLP, the approaches of Brodka et al. and Greene et al. can track communities very well.

For the dataset AS, shown in Fig~\ref{fig:o_c} and Fig~\ref{fig:o_f}, we can draw a similar conclusion. Here it should be noted that the approach of Takaffoli et al. can track certain communities very well where other approaches fail, but the broad color range indicates that the performance of this approach is not stable enough.

Overall, the approach of Brodka et al. and approach of Greene et al. is able to track a community very well in terms of tracking quality on datasets which have overlapping communities.\\

\subsection{Experiment Based on Non-overlapping Communities Extracted by Infomap}

For the second experiment, the group detection algorithm Infomap was used on each snapshot. Table~\ref{tab:infomap} contains the description of the processed data: The first column gives the average number of communities extracted by Infomap per snapshot, and the second shows the average size of the group.

\begin{table}[t]
    \centering
    \caption{Data Descriptions for Disjoint Communities.}
    \begin{tabular}{{@{}ccc@{}}}
        \toprule
         Network    & \#avg\_com      &   \#com\_size    \\\midrule
         DBLP       &     595        &       11          \\
         AS         &     129           &     17           \\
         YELP       &    222            &     16　         \\\bottomrule
    \end{tabular}
    \label{tab:infomap}
\end{table}

Table~\ref{tab: threshold_d} gives the similarity threshold values from each of the tracking methods on disjoint communities.\\

\begin{table}[t]
\centering
\caption{Similarity Threshold for Each Method on Disjoint Communities. Greene et al. using Jaccard coefficient, Takaffoli et al. using Modec similarity, Brodka et al. using the Inclusion similarity and Tajeuna et al. Mutual transition.}
\label{tab: threshold_d}
\begin{tabular}{@{}cccccc@{}}
\toprule
     & Greene et al.  & Takaffoli et al.  & Brodka et al.  & Tajeuna et al.  \\ \midrule

DBLP &      0.11      &       0.15      &     0.10 / 0.11   &   0.15   \\\midrule
AS   &     0.42      &    0.44      &     0.43 / 0.41       &        0.30               \\\midrule
YELP &      0.04         &         0.06               &        0.03 / 0.03                 &     0.08                  \\\bottomrule
\end{tabular}
\end{table}

\subsubsection{Quantity of Tracking}
\label{quantity_d}

\begin{figure}[!ht]
	\centering
    \caption{Tracking Quantities on Overlapping Communities.}
    \label{fig: quantity_d}
    \includegraphics[width=0.5\textwidth]{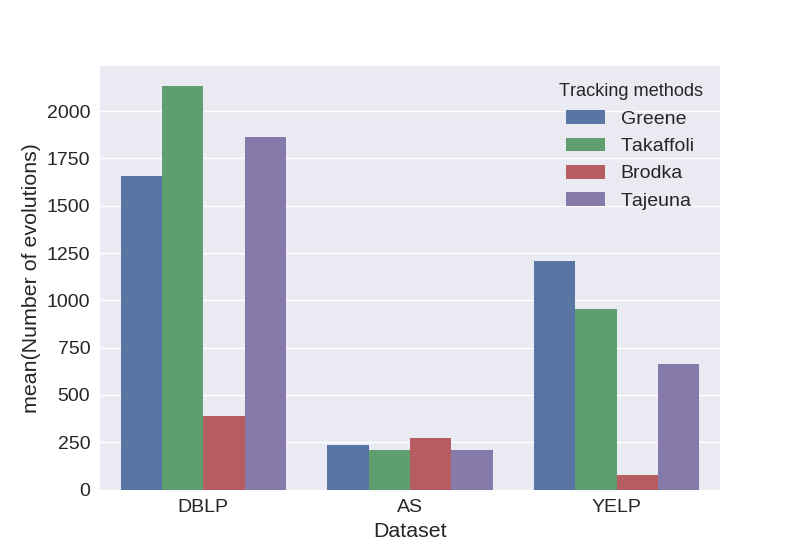}
\end{figure}

From the numbers of evolving communities found by different approaches on each dataset, shown in Fig~\ref{fig: quantity_d}, we obtain observations similar to those in Section~\ref{quatity_o}. Brodka’s approach has more chance of receiving a good audience in tracking only big-size communities when the quantity of tracking is of interest, and the approaches of Takaffoli et al., approach of Tajeuna et al. and Greene et al. are efficient enough to capture most potential evolutions over time.\\

\begin{figure*}[!ht]
    \centering
    \subfloat[YELP\_APCC.]{\includegraphics[width=0.3\linewidth]{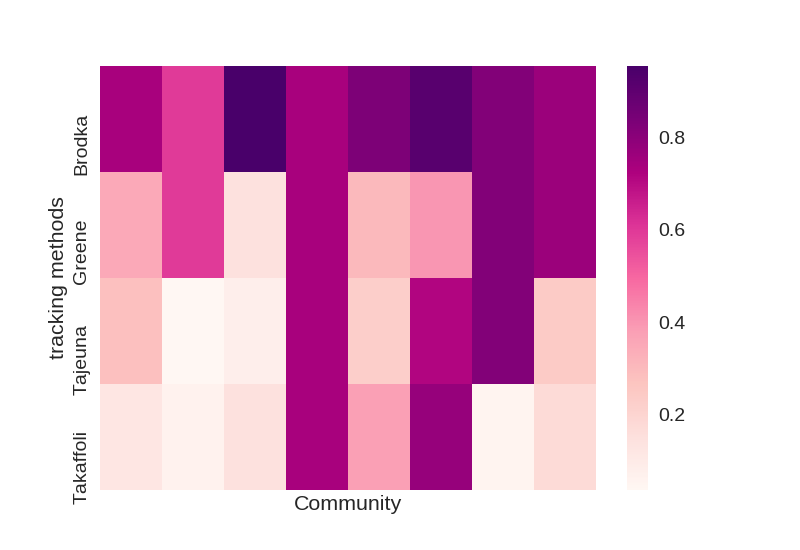}\label{fig:d_a}}\quad
    \subfloat[DBLP\_APCC.]{\includegraphics[width=0.3\linewidth]{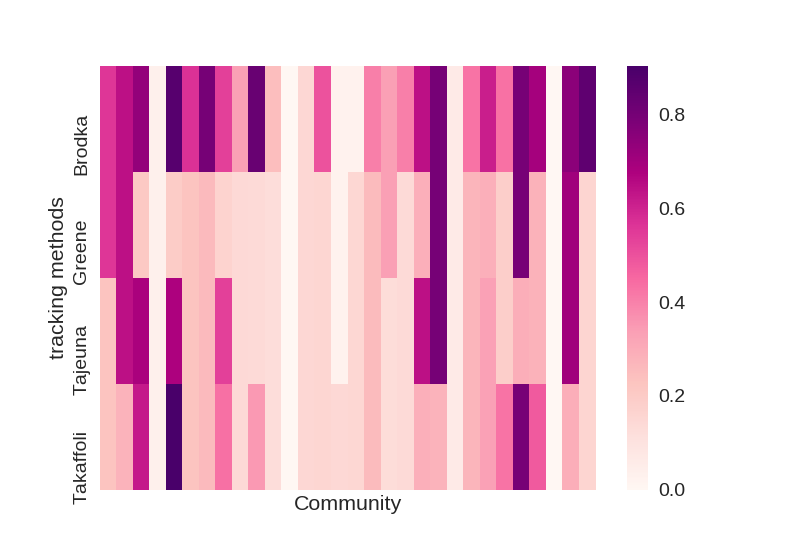}\label{fig:d_b}}\quad
    \subfloat[AS\_APCC.]{\includegraphics[width=0.3\linewidth]{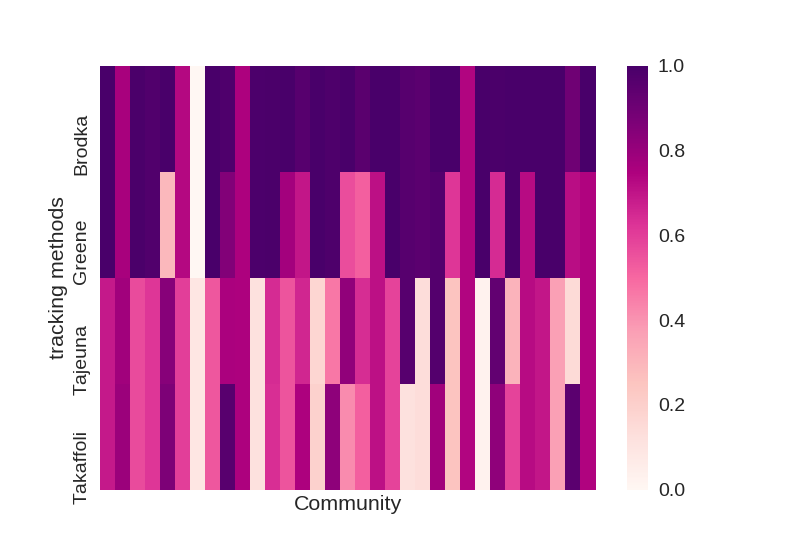}\label{fig:d_c}}\quad
    \medskip
ou    \subfloat[YELP\_APNP.]{\includegraphics[width=0.3\linewidth]{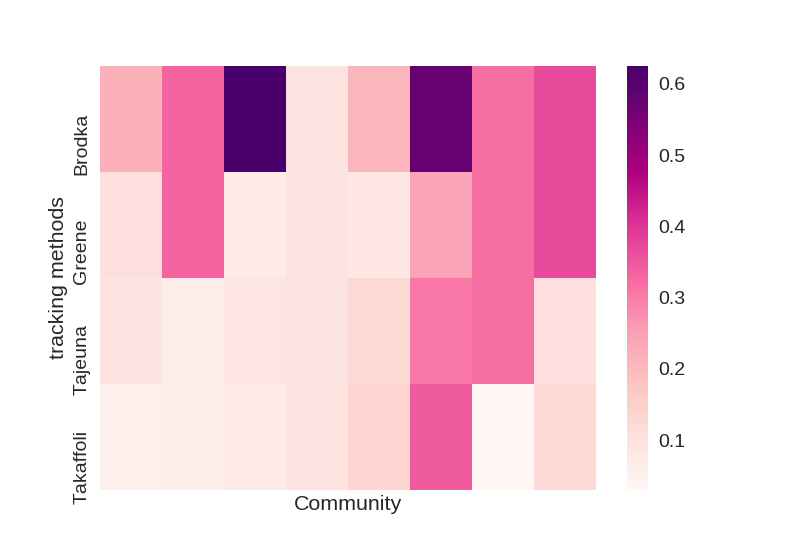}\label{fig:d_d}}\quad
    \subfloat[DBLP\_APNP.]{\includegraphics[width=0.3\linewidth]{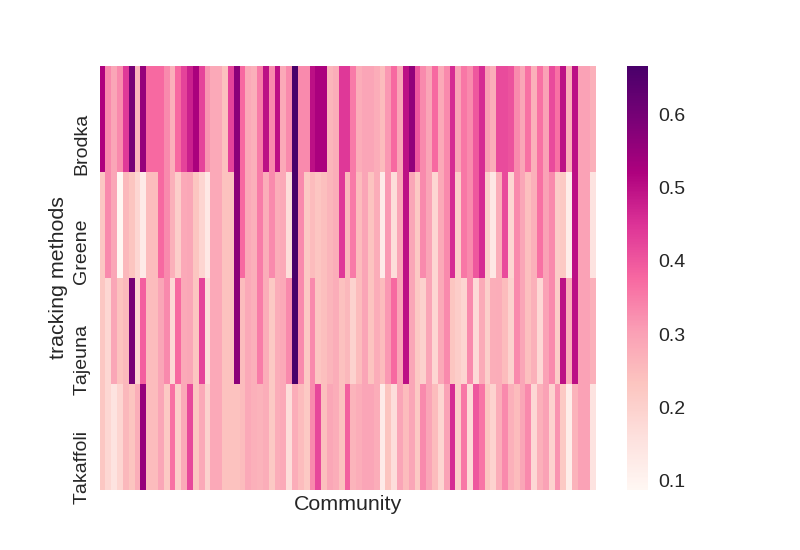}\label{fig:d_e}}\quad
    \subfloat[AS\_APNP.]{\includegraphics[width=0.3\linewidth]{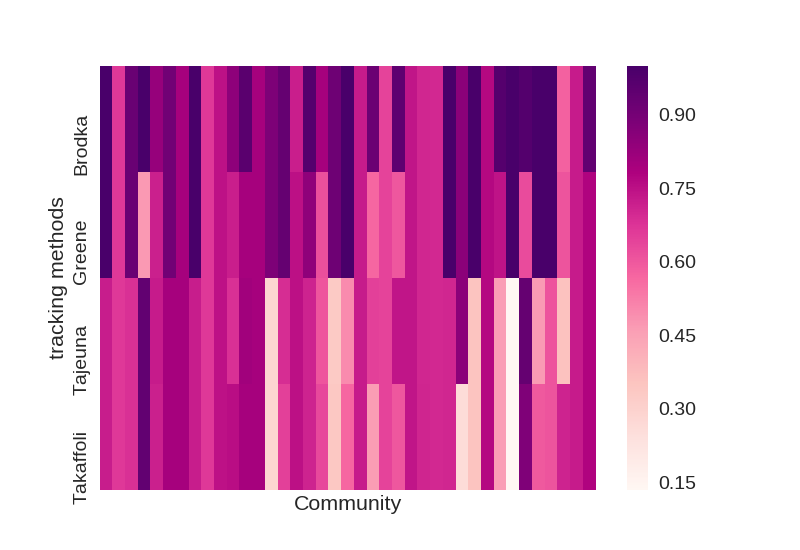}\label{fig:d_f}}
    \caption{In this heatmap, the darker the color of the cell, the higher the value. For (b) and (e) we removed the communities which all four communities track well (APCC and APNP are all above $0.3$) and for (c) and (f) we again removed the communities which all four approaches track well (APCC and APNP are all above $0.8$).}
\label{fig: evaluation_d}
\end{figure*}

\subsubsection{Quality of Tracking}
After implementing the four tracking methods on the three non-overlapping datasets, the community sequences obtained were processed by the same method used on overlapping communities.
Fig~\ref{fig: evaluation_d} is the heatmap of APCC and APNP. The interpretation of the heatmap has already been given in Sec~\ref{quality_o}.

The numbers of communities selected as starting points for an evolution which all the approaches can successfully track, used to plot the heatmap, are as follows: 
\begin{itemize}
\item For DBLP, there are 332 communities.
\item For AS, there are 170 communities.
\item For YELP, there are 8 communities.
\end{itemize}

Fig~\ref{fig: evaluation_d} contains the heatmap for APCC and APNP. In Fig~\ref{fig:d_a}, Fig~\ref{fig:d_b} and Fig~\ref{fig:d_c}, each column gives the APCC for the evolutions of the same community tracked by the approaches of Brodka et al., Greene et al., Tajeuna et al. and Takaffoli et al.. Fig~\ref{fig:d_d} Fig~\ref{fig:d_e} and Fig~\ref{fig:d_f} give the same illustration for the APNP.


Again, similar conclusions can be reached. In all the sub-figures of Fig~\ref{fig: evaluation_d}, most of the darker cells are located in the row corresponding to the approach of Brodka et al., and that of Greene et al. and Takaffoli et al. can be very good, according to the broad color range. We conclude that the approach of Brodka et al. can also track a community very well on datasets which have non-overlapping communities.\\



\section{Conclusion}
In this paper, we have presented a comparative study of approaches for tracking communities in time-evolving social networks in which communities may be either overlapping or disjoint. We introduced six existing popular tracking algorithms and tested four of them in our Experiment section (Section~\ref{section: first}).\\

For purposes of evaluation, we introduced two measures, the Average Pearson Correlation and the Proportion of Nodes Persisting, to evaluate the quality of the communities tracked. We performed the comparison study on real data from the DBLP, AS and YELP datasets (note that in our case, we consider the communities detected in dataset AS to be large-size communities, those from YELP to be medium-size, and those from DBLP to be small-size; for details see Table~\ref{tab:cpm} and Table~\ref{tab:infomap}). 

Our comparison reveals that all the approaches are capable of tracking overlapping and disjoint communities over time into sequences in which the set of communities shows good global resemblance (above the threshold of similarity). Generally, looking at both overlapping and disjoint communities, we observed that all the approaches are capable of tracking community evolutions very well. Normally approaches of Greene et al., Takaffoli et al. and Tajeuna et al. are capable of tracking a certain number of evolutions of communities over time. Approach of Brodka et al. can only find sufficient number of evolutions when tracking big-size communities, but achieve especially satisfying results with respect to APCC and APNP values on most dataset (in our experiment we only demonstrated 3 representative real datasets, so here we cannot popularize it on all datasets).

We conclude that when the dataset which will be dealt with has big-size communities, or, quality of tracking is focused, approach of Brodka et al. is the best choice, if the quantity of tracking is of interest, approaches of Greene et al., Takaffoli et al. and Tajeuna et al. are worth trying. 

In the future, we will focus on tracking and predicting critical events a community may undergo.

\balance
\bibliographystyle{IEEEtran}
\bibliography{IEEEabrv,bare_conf_revised.bib}

\begin{thebibliography}{10}
\providecommand{\url}[1]{#1}
\csname url@samestyle\endcsname
\providecommand{\newblock}{\relax}
\providecommand{\bibinfo}[2]{#2}
\providecommand{\BIBentrySTDinterwordspacing}{\spaceskip=0pt\relax}
\providecommand{\BIBentryALTinterwordstretchfactor}{4}
\providecommand{\BIBentryALTinterwordspacing}{\spaceskip=\fontdimen2\font plus
\BIBentryALTinterwordstretchfactor\fontdimen3\font minus
  \fontdimen4\font\relax}
\providecommand{\BIBforeignlanguage}[2]{{%
\expandafter\ifx\csname l@#1\endcsname\relax
\typeout{** WARNING: IEEEtran.bst: No hyphenation pattern has been}%
\typeout{** loaded for the language `#1'. Using the pattern for}%
\typeout{** the default language instead.}%
\else
\language=\csname l@#1\endcsname
\fi
#2}}
\providecommand{\BIBdecl}{\relax}
\BIBdecl

\bibitem{lancichinetti2008benchmark}
A.~Lancichinetti, S.~Fortunato, and F.~Radicchi, ``Benchmark graphs for testing
  community detection algorithms,'' \emph{Physical review E}, vol.~78, no.~4,
  p. 046110, 2008.

\bibitem{leskovec2010empirical}
J.~Leskovec, K.~J. Lang, and M.~Mahoney, ``Empirical comparison of algorithms
  for network community detection,'' \textit{In Proceedings of the 19th
  International Conference on World Wide Web (WWW)}, pp. 631--640, 2010.

\bibitem{sarkar2005dynamic}
P.~Sarkar and A.~W. Moore, ``Dynamic social network analysis using latent space
  models,'' \emph{ACM SIGKDD Explorations Newsletter}, vol.~7, no.~2, pp.
  31--40, 2005.

\bibitem{takaffoli2011tracking}
M.~Takaffoli, J.~Fagnan, F.~Sangi, and O.~R. Za{\"\i}ane, ``Tracking changes in
  dynamic information networks,'' \textit{International Conference on
  Computational Aspects of Social Networks (CASoN)}, pp. 94--101, 2011.

\bibitem{asur2009event}
S.~Asur, S.~Parthasarathy, and D.~Ucar, ``An event-based framework for
  characterizing the evolutionary behavior of interaction graphs,'' \emph{ACM
  Transactions on Knowledge Discovery from Data (TKDD)}, vol.~3, no.~4, p.~16,
  2009.

\bibitem{lee2014incremental}
P.~Lee, L.~V. Lakshmanan, and E.~E. Milios, ``Incremental cluster evolution
  tracking from highly dynamic network data,'' \textit{30th International
  Conference on Data Engineering (ICDE)}, pp. 3--14, 2014.

\bibitem{tajeuna2015tracking}
E.~G. Tajeuna, M.~Bouguessa, and S.~Wang, ``Tracking the evolution of community
  structures in time-evolving social networks,'' \textit{International
  Conference on Data Science and Advanced Analytics (DSAA)}, pp. 1--10, 2015.

\bibitem{bhat2015hoctracker}
S.~Y. Bhat and M.~Abulaish, ``Hoctracker: Tracking the evolution of
  hierarchical and overlapping communities in dynamic social networks,''
  \emph{IEEE Transactions on Knowledge and Data engineering}, vol.~27, no.~4,
  pp. 1019--1013, 2015.

\bibitem{rossi2013modeling}
R.~A. Rossi, B.~Gallagher, J.~Neville, and K.~Henderson, ``Modeling dynamic
  behavior in large evolving graphs,'' \textit{In proceedings of the 6th
  International Conference on Web Search and Data Mining}, pp. 667--676, 2013.

\bibitem{skillicorn2013spectral}
D.~B. Skillicorn, Q.~Zheng, and C.~Morselli, ``Spectral embedding for dynamic
  social networks,'' \textit{In proceedings of the 2013 IEEE/ACM International
  Conference on Advances in Social Networks Analysis and Mining}, pp. 316--323,
  2013.

\bibitem{calvo2004social}
A.~Calv{\'o}-Armengol and Y.~Zenou, ``Social networks and crime decisions: The
  role of social structure in facilitating delinquent behavior,''
  \emph{International Economic Review}, vol.~45, no.~3, pp. 939--958, 2004.

\bibitem{luke2007network}
D.~A. Luke and J.~K. Harris, ``Network analysis in public health: history,
  methods, and applications,'' \emph{Annu. Rev. Public Health}, vol.~28, pp.
  69--93, 2007.

\bibitem{greene2010tracking}
D.~Greene, D.~Doyle, and P.~Cunningham, ``Tracking the evolution of communities
  in dynamic social networks,'' \textit{International Conference on Advances in
  social networks analysis and mining (ASONAM)}, pp. 176--183, 2010.

\bibitem{brodka2013ged}
P.~Br{\'o}dka, S.~Saganowski, and P.~Kazienko, ``Ged: the method for group
  evolution discovery in social networks,'' \emph{Social Network Analysis and
  Mining}, vol.~3, no.~1, pp. 1--14, 2013.

\bibitem{gliwa2013different}
B.~Gliwa, P.~Br{\'o}dka, A.~Zygmunt, S.~Saganowski, P.~Kazienko, and J.~Kozlak,
  ``Different approaches to community evolution prediction in blogosphere,''
  \textit{In proceedings of the 2013 IEEE/ACM International Conference on
  Advances in Social Networks Analysis and Mining (ASONAM)}, pp. 1291--1298,
  2013.

\bibitem{tang2014detecting}
X.~Tang and C.~C. Yang, ``Detecting social media hidden communities using
  dynamic stochastic blockmodel with temporal dirichlet process,'' \emph{ACM
  Transactions on Intelligent Systems and Technology (TIST)}, vol.~5, no.~2,
  p.~36, 2014.

\bibitem{blondel2008fast}
V.~D. Blondel, J.-L. Guillaume, R.~Lambiotte, and E.~Lefebvre, ``Fast unfolding
  of communities in large networks,'' \emph{Journal of statistical mechanics:
  theory and experiment}, vol. 2008, no.~10, p. P10008, 2008.

\bibitem{chen2009local}
J.~Chen, O.~Za{\"\i}ane, and R.~Goebel, ``Local community identification in
  social networks,'' \textit{International Conference on Advances in Social
  Network Analysis and Mining (ASONAM)}, pp. 237--242, 2009.

\bibitem{adamcsek2006cfinder}
B.~Adamcsek, G.~Palla, I.~J. Farkas, I.~Der{\'e}nyi, and T.~Vicsek, ``Cfinder:
  locating cliques and overlapping modules in biological networks,''
  \emph{Bioinformatics}, vol.~22, no.~8, pp. 1021--1023, 2006.

\bibitem{rosvall2008maps}
M.~Rosvall and C.~T. Bergstrom, ``Maps of random walks on complex networks
  reveal community structure,'' \emph{Proceedings of the National Academy of
  Sciences}, vol. 105, no.~4, pp. 1118--1123, 2008.

\bibitem{fortunato2010community}
S.~Fortunato, ``Community detection in graphs,'' \emph{Physics reports}, vol.
  486, no.~3, pp. 75--174, 2010.

\bibitem{jaccard1912distribution}
P.~Jaccard, ``The distribution of the flora in the alpine zone.'' \emph{New
  phytologist}, vol.~11, no.~2, pp. 37--50, 1912.

\end{thebibliography}
%

\end{document}